\documentclass[journal]{vgtc}                
\ifpdf
  \pdfoutput=1\relax                   
  \usepackage{graphicx}                
  \DeclareGraphicsExtensions{.pdf,.png,.jpg,.jpeg} 
\else
  \usepackage{graphicx}                
  \DeclareGraphicsExtensions{.eps}     
\fi%

\graphicspath{{figures/}{pictures/}{images/}{./}} 

\usepackage{microtype}                 
\usepackage{times}                     
\usepackage{cite}                      
\usepackage{booktabs}                  
\usepackage[strings]{underscore}
\usepackage{tabularx}
\usepackage{multirow}
\usepackage{fontawesome5}
\usepackage{xcolor}
\usepackage{siunitx}
\usepackage{upgreek}
\usepackage[normalem]{ulem} 
\usepackage{capt-of}
\usepackage{siunitx}
\usepackage{float}
\usepackage{wrapfig}
\usepackage{enumitem}

\definecolor{myGreen}{HTML}{42bc82}

\onlineid{0}

\vgtccategory{Research}
\vgtcpapertype{evaluation}

\title{Characterizing Visualization Insights through\\ Entity-Based Interaction: An Exploratory Study}

\author{Chen He, Tung Vuong, and Giulio Jacucci}
\authorfooter{
\item
 Chen He, Tung Vuong and Giulio Jacucci are with the Department of Computer Science, University of Helsinki. E-mail: chen.he@helsinki.fi, vuong@cs.helsinki.fi, and giulio.jacucci@helsinki.fi.
}

\shortauthortitle{He \MakeLowercase{\textit{et al.}}: Characterizing Visualization Insights}

\abstract{
One of the primary purposes of visualization is to assist users in discovering insights. While there has been much research in information visualization aiming at complex data transformation and novel presentation techniques, relatively little has been done to understand  
how users derive insights through interactive visualization of data.
This paper presents a crowdsourced study with 158 participants investigating the relation between entity-based interaction (an action + its target entity) and the resulting insight.
To this end, we generalized the interaction with an existing CO\textsubscript{2} Explorer as entity-based interaction and enabled users to input notes and refer to relevant entities to assist their narratives. We logged interactions of users freely exploring the visualization and characterized their externalized insights about the data.
Using entity-based interactions and references to infer insight characteristics (category, overview versus detail, and prior knowledge), we found evidence that compared with interactions, entity references improved insight characterization from slight/fair to fair/moderate agreements. To interpret prediction outcomes, feature importance and correlation analysis indicated that, e.g., detailed insights tended to have more mouse-overs in the chart area and cite the vertical reference lines in the line chart as evidence. 
We discuss study limitations and implications on knowledge-assisted visualization, e.g., insight recommendations based on user exploration.
} 


\vgtcinsertpkg

\begin{document}


\maketitle
\raggedbottom

\section{Introduction} \label{sec:introduction}
Interactive visualization allows users to interact with the visual depiction of the data to gain insight into the data \cite{visexplore, infovisMining}. For this purpose, researchers identified insight generation as a primary goal of visualization \cite{understanding, ism}. 
To evaluate how well a visualization supports insight, Saraiya et al. \cite{insight} proposed an insight-based methodology in which they considered an insight as an individual observation of the data and measured the characteristics of generated insights, such as insight category and breadth versus depth, to inform the strengths and weaknesses of visualization in supporting different types of insights. 
However, to understand visualization insight, we cannot overlook the role of interaction. User interaction results from user intent, reflects user reasoning and sense-making process, and is directly related to the resulting insight \cite{science}. Through correlating interaction types and insight characteristics, prior work revealed the tight connection between interaction and insight. For instance, studies showed that exploration actions promoted the number of insights \cite{caseStudy} and could lead to unexpected discoveries about the data \cite{he}. These findings imply the potential of automatically characterizing insight using interactions, alleviating user efforts in tagging insights \cite{commentspace} while facilitating insight management and recommendation \cite{click2annotate}. 

Usually, a user interaction could be identified as an action and its target entity. For instance, on a map, users could zoom in (action) on a district (entity) to explore the locations of facilities. 
Generalizing user interaction with visualization as entity-based interaction allows users to maintain the focus of the interaction and aspects of the origin of insights. As an example, individual lines in a line chart could be taken as interaction targets and insight indicators; users can mouse over the line to view data details and link the line with a text description of their discovery. As the concept of entity-based interaction commonly exists in visualization, in this study, we identified all user interactions with an existing visualization---CO\textsubscript{2} Explorer---as entity-based interactions, such as \textit{select a year} and \textit{mouse-over a country}, and allowed users to input notes and cite relevant entities to assist their narratives. 

We studied the Explorer through crowdsourcing following the insight-based methodology \cite{insight}, allowing open-ended user exploration and insight discovery, to explore the research question (RQ): \textit{To which extent can entity-based interactions/references infer insight characteristics?} For instance, could we characterize insights as detailed rather than overviews if users elaborate on and refer to individual data points for the discovered insight?
To study the RQ, we recorded user interactions and insights of exploring the CO\textsubscript{2} Explorer and used the number of different types of interactions and entity references as predictors to infer insight characteristics of insight category, overview versus detail, and using prior knowledge with two advanced machine learning models. To interpret model outcomes, we present feature importance on individual cases and participant-wise correlation analysis between insight characteristics and the predictors. 

The main contribution of this research is two-fold.
First, we present an approach to characterize insight using entity-based interactions and references. We identify interaction targets as entities to be cited to support insight narratives. 
Second, we demonstrate experimental evidence that compared with using entity-based interactions, using entity references improved insight characterization agreement from fair to moderate on insight categories, and from slight to fair on overview versus detail. Interpretations showed several interesting tendencies. For example, users tended to cite vertical reference lines in the line chart to assist the narration of comparison and detailed types of insights.

Study results imply the potential of automatically characterizing insights to, e.g., provide recommendations based on user exploration. Following the above example, visualizations could recommend other diverse insights if the user constantly refer to vertical reference lines.
A comparison with previous discoveries (e.g., \cite{selfer, he}) revealed that interactions / entity references could infer insight characteristics, but the particular entity and insight relations could be relevant to the type of dataset and visualization being studied. 
Nonetheless, the approach we used to explore the RQ is applicable to study other visualizations, as explained in \autoref{sec:conclusion}, which discusses implications on knowledge-assisted visualization and study limitations.

\section{Related work}
This section reviews how this work advances existing research on studying visualization insight and interactions and informs the selections of insight characteristics for the study.

\subsection{Insight-based evaluation}
Researchers leveraged the insight-based evaluation extensively to assess visualization design (e.g., \cite{style, latency, selfer}). Generally, the insight-based method is open-ended, putting no constraints on the insights users can report or the time users can spend; then the collected insights are coded, such as categorized, by two or more evaluators to assess the visualization in supporting different types of insights \cite{toward}. 
Smuc et al. further proposed to assess how insights build on one another or prior knowledge for visualization to better support the discovery process \cite{score}. However, analyzing insight alone can not uncover how users discover insight \cite{modeling}. 

To address this issue, researchers proposed analyzing insight together with user interaction logs to gain a more holistic view of the data exploration process \cite{modeling, rome, caseStudy}. For instance, Reda et al. \cite{modeling} proposed drawing transition graphs of user interaction and mental states collected through interaction logs and think-aloud protocol to capture the exploration process a visualization tool affords. 
Guo et al. \cite{caseStudy} correlated the number of interactions and insights of users using a visualization tool and discovered that exploration actions promoted insight generation whereas filtering actions inhibited it.
Further, He et al. \cite{he} analyzed an insight from multiple perspectives and correlated the scale of insight characteristics to the number of interactions by studying a domain-related visualization. They found evidence that exploration actions related to unexpected insight, and the pattern of drilling down into data of interest led to insight with high domain values.

Building on the promising results from prior studies, this research aims to infer insight characteristics using interactions through machine learning to shed light on knowledge-assisted visualization in which users share insight during visual exploration. For instance, visualization tools could recommend insight or guide insight discovery based on user interaction.

\subsection{Interaction analysis with entities}
Interaction contains rich user information and can be used for provenance, visualization evaluation, reasoning \& sensemaking, and prediction \& recommendation \cite{he2022entity}. Blascheck et al. \cite{va2, data-rich} visually integrated user interaction, eye tracking, and think-aloud data to support the analysis of user data exploration process.
Researchers also visualized interactions to allow users to recover their own / others' reasoning processes \cite{recovering, SensePath, helping}. Moreover, interactions were analyzed to help guide users along the analytic process \cite{grammar}, predict personality traits \cite{waldo}, support the recommendation of relevant resources for further exploration \cite{focus, semantic}, etc. As a primary goal of visualization is to support insight, we attempt to explicitly relate interactions to the resulting insight to help better understand insight generation.

For analysis, interaction is characterized at multiple levels of granularity as tasks, sub-tasks, actions, and events \cite{characterizing}. The action level, generic toward the domain yet semantically meaningful, is often captured for insight provenance \cite{characterizing, caseStudy, he} and intent modeling \cite{I}.
Entity-based interaction also facilitates from the action level to categorize interactions.  

Entities are widely used to represent any real-world objects and concepts. In information visualization, combinations of entities can provide an overview of the information space; entities can be organized as one's mental representation of the information space and shared to support collaboration \cite{hypercues}. For instance, PivotPaths visualizes three types of entities including people, resources, and concepts in node-link diagrams and allows pivot actions to explore entity relations \cite{pivotpaths}. QueryTogether enables users to share entities to support multi-device co-located collaborative search \cite{querytogether}.
In this research, we identify interactions as actions plus their target entities. This allows us to further capture these entities for references to assist user insight narratives which we elaborate on in \autoref{sec:entityInteraction}. 

\subsection{Insight characterization} \label{relatedWork:chara}
Characterizing insight in prior work primarily aimed at visualization evaluation, insight management, and supporting automation. 
To evaluate visualization design, Saraiya et al. characterized insight by the number of observations, hypothesis, time to reach insight, category, directed versus unexpected, breadth versus depth, correctness, and domain value \cite{insight}. This characterization, albeit derived from a domain-relevant perspective, has been applied in many other studies (e.g., \cite{modeling, he, style}). To support insight organization, related work allowed users to tag comments as replies, suggestions, links, etc. by Bunt et al. \cite{taggedComments}, and as hypotheses, questions, to-dos, evidence for, and evidence against by Willett et al. \cite{commentspace}.
With a different purpose of supporting automation, researchers categorized insight in a bottom-up manner of people reading static charts to build systems that generate the most suitable visualization for a particular type of insight \cite{composing, selfer}.
Law et al. \cite{dataInsight} provided a theoretical characterization of insight through interviewing professional visualization users.

Compared with Law et al.'s characterization \cite{dataInsight}, we adopted Saraiya et al.'s approach to characterize insight considering its practicality in empirical studies; we used the characteristics that could be applied to generic visualizations. 
TaggedComments \cite{taggedComments} and CommentSpace \cite{commentspace} characterized insight in a concrete way, which to our opinion could be dealt with using natural language processing, whereas Saraiya et al.'s characterization is more abstract, which could benefit from the analysis of interactions and referred entities for automation, though both types of characterization are meaningful for insight management. Coupled with the finding from prior studies that the use of domain knowledge affect insight generation \cite{he, score}, we characterized insights by category, overview versus detail, using prior knowledge, and correctness.

\begin{table*}[!t]
  	\includegraphics[width=\textwidth]{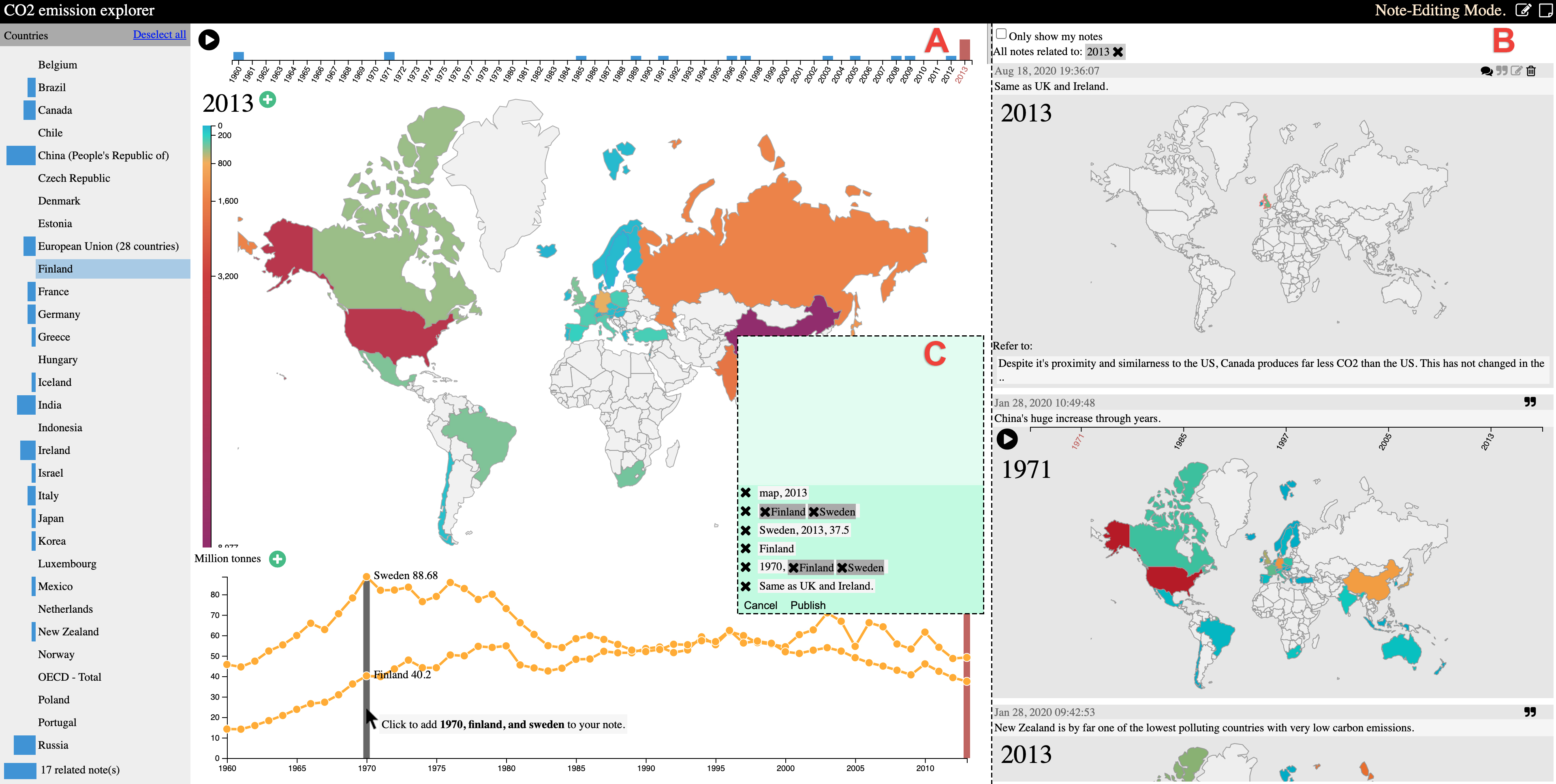}
	\centering
	\vspace{-6mm}
    \captionof{figure}{Interface of the CO\textsubscript{2} Explorer. Users can select a year from the top list to view that year's global emissions in the choropleth map and select countries from the left list to view their historical emissions in the line chart (A). The note column (B) shows notes related to the filtered year 2013. Users can refer to six types of entities to support insight externalization (\autoref{tab:visualEntity}); their citations are shown in the sticky note (C). A tooltip is shown on hovering to instruct users to cite visual entities. Black dashed lines and red capital letters in the figure are added for demonstration purposes.}
    \label{fig:interface}
    \vspace{3mm}
    \captionof{table}{Types of entities, how to cite them, and their contributions to the result view.}
    \vspace{-1mm}
\begin{tabularx}{\textwidth}{p{.01\textwidth} | >{\raggedleft}p{.1\textwidth} p{.82\textwidth} } 
\toprule
    \multicolumn{2}{r}{Entity type} & How to cite and the contribution to the result view \\
    \midrule
    \multirow{2}{*}{\rotatebox{90}{\parbox{2.5cm}{\centering Chart}}}
    &Map & Cite a map by clicking on the corresponding \textcolor{myGreen}{\faIcon{plus-circle}} icon. The first row with ``map, 2013'' in \autoref{fig:interface} (C) indicates a map citation. If users add multiple maps, the result view will show a year list, similar to the one in the visualization, to support navigation. \\
    &Line chart & Cite a line chart in the same way as citing a map. The second row with ``\faIcon{times}Finland  \faIcon{times}Sweden'' exemplifies a line chart citation. Users can remove individual countries by clicking on their \faIcon{times} icons if the countries are irrelevant to the note. In case of multiple line chart references, the result view generates one line chart integrating the data of all referred countries.\\
    \midrule
    \multirow{2}{*}{\rotatebox{90}{\parbox{2.5cm}{\centering Chart component}}}&
    Map point & Cite a country on the map by clicking on the country. The row of ``Sweden, 2013, 37.5'' denotes a referred map point. Referring to map points from various years generates a year list in the result view similar to the map citations.\\
    &Line & Cite a line from the line chart by clicking on the line. The row with ``Finland'' is a line citation.\\
    &Vertical reference line & Cite the data of a specific year in the line chart by clicking on the black vertical line (\autoref{fig:interface}). The row of ``1970, \faIcon{times}Finland \faIcon{times}Sweden'' represents this type of citation. Similar to the line chart citation, users can remove irrelevant countries to make the reference more focused. The result view shows a black vertical reference line as well as the lines of the countries in the line chart.\\
    \midrule
    &Note & Cite a note by clicking on the \faIcon{quote-right} icon of the corresponding note. The last row in the sticky note represents a note citation. No limited numbers of notes users can cite to create discussions.\\
\bottomrule
\end{tabularx}
\vspace{-3mm}
\label{tab:visualEntity}
\end{table*}

\section{The CO\textsubscript{2} Explorer} \label{sec:design}

We used a CO\textsubscript{2} Explorer, previously studied by Boy et al. \cite{storytelling} and Feng et al. \cite{hindsight}, for this research, facilitating result comparison across multiple studies.
The Explorer consists of a map view and a line chart (\autoref{fig:interface} (A)). Users can select a year from the top list to view that year's global CO\textsubscript{2} emission on the choropleth map or play the year list to view the animation of the yearly global emission maps.
Mousing over a country on the map displays a tooltip with the country name and its emission value. Users can also select countries from the left list to view these countries' histories of CO\textsubscript{2} emission in the line chart. Mousing over the line chart displays a black vertical reference line marking the year nearest to the mouse location and that year's emission values of selected countries. The red vertical line in the line chart indicates the selected year in the map view. Users can deselect all selected countries by clicking on the button at the top of the country list. The dataset is from an online library that contains the annual CO\textsubscript{2} emission data of 43 countries from 1960 to 2013 \cite{co2}.

\subsection{Entity-based interaction} \label{sec:entityInteraction}
After we generalize the interactions to entity-based actions, the Explorer supports
12 actions:
\begin{itemize}[leftmargin = *]
    \setlength\itemsep{-.3mm}
    \item Interact with the left country list: \textit{select a country; deselect a country; deselect all selected countries; mouse-over a country.}
    \item  Interact with the top year list: \textit{select a year; mouse-over a year; play; stop.}
    \item Interact with the map: \textit{mouse-over a map point.} 
    \item Interact with the line chart: \textit{mouse-over a line; mouse-over the area of a specific year; mouse-over a vertical reference line.}
\end{itemize}

\subsubsection{Entity references to support insight}\label{prototype:entity}
We suggest incorporating insight externalization as an integral part of visual exploration under the umbrella of supporting insight. Prior work identified various entities in a visualization that could be annotated \cite{ChartAccent, entity}. ChartAccent distinguished targets in a chart for annotation, including data items, chart elements, coordinate spaces, and prior annotations \cite{ChartAccent}, whereas Lai et al. \cite{entity} categorized two types of entities in a visualization: a data entity is a visual element that encodes data, such as a bar in a bar chart, whereas an auxiliary entity assists visual reading, such as a color legend and a data label. 

In this study, the visual entities identified earlier as interaction targets prompted us to consider two categories of entities for insight, ranging from a holistic visualization consisting of multiple charts to a basic visual element: 
a chart-level entity represents various charts, such as a scatter plot and a parallel coordinates plot, whereas a chart component-level entity depicts a chart component, such as a cluster or a point in a scatter plot and lines between three dimensions in parallel coordinates. With the Explorer, chart-level entities are maps and line charts, whereas chart component-level entities are lines and vertical reference lines in the line chart and individual countries on the map.

In contrast to ChartAccent enabling users to attach text input to target entities, this study allowed users to attach entities to the notes. This enabled users to consider data as references to their expression while describing a cohesive story as an insight, such as involving external information and prior knowledge; meanwhile, it facilitated insight management: by recognizing data dimensions in the referred visual entities, the Explorer can support scented insight browsing \cite{click2annotate, scentedWidgets}. Additionally, same as ChartAccent, we enabled users to refer to a prior insight to construct a unified mental model of exploring visualization and insight.

\textit{Edit notes.} After we apply entity references in the Explorer, there are two ways to start composing a note: by clicking on the \faIcon{edit} icon at the top right corner or by replying to a note through clicking on its \faIcon{quote-right} icon. In the latter case, the clicked note entity is added as a reference in the current note. Besides citing notes, users can refer to five types of visual entities to support externalization, as mentioned earlier. During note editing, hovering over the visual elements, users can view instructions on referring to a certain entity in a tooltip (\autoref{fig:interface}). After users publish a note, a result view is generated based on the referred entities. \autoref{tab:visualEntity} explains how to cite the various types of entities and their contributions to the result view. Possible entity-based editing actions with the Explorer are:
\begin{itemize}[leftmargin = *]
    \setlength\itemsep{-.3mm}
\item Enter or exit note editing: \textit{open a note input; reply to a note; save a note; cancel a note input; open a note editing; update a note; delete a note.}
\item During editing: \textit{mouse-over the text area; mouse-over a referred entity in the sticky note; add an entity; remove an entity; remove a country from the vertical line reference; remove a country from the line chart reference.}
\end{itemize}

\textit{Explore notes.} 
The Explorer enables users to view others' published notes by clicking on the \faIcon[regular]{sticky-note} icon at the top right corner to show notes. To support scented insight browsing \cite{scentedWidgets, click2annotate}, the Explorer extracts countries and years referred in notes and attaches the counts of relevant notes to the country and year lists as blue bars. Users can click on the bar to filter notes. The filtered notes are displayed in the note column in reverse chronological order (\autoref{fig:interface} (B)). 
The checkbox at the top of the note column allows users to hide others' notes.
To explore discussions, users click on the \faIcon{comments} icon of a note. The note column then displays at most 20 discussion notes to the selected note in chronological order. Meanwhile, a graph depicting the discussion flow is shown to the right side. Each node in the graph represents a note; links point at the flow of the discussion. A red node indicates the selected note. Discussion notes and the graph are coordinated, as mousing over a node/note triggers a dark yellow highlight for the focused note/node and a lighter yellow highlight for related notes/nodes.
Clicking on a node can trigger the note column scrolling to the corresponding note if the note is outside the view. There are 22 entity-based note exploration actions:
\begin{itemize}[leftmargin = *]
    \setlength\itemsep{-.3mm}
\item Control the note display: \textit{show notes; hide notes; only show my notes; also show public notes.} 
\item Scented note browsing: \textit{view notes of a country; view notes of a year; remove the note filter.} 
\item Interact with the content of a note: \textit{mouse-over texts of a note; mouse-over a referred note in a note.} 
\item Interact with the attached visualization of a note (similar to the data exploration actions): \textit{select a year; mouse-over a year; play; stop; mouse-over a map point; mouse-over a line; mouse-over the area of a specific year; mouse-over a vertical reference line.} 
\item Explore discussions: \textit{view discussions; remove the discussion filter; drag a node; mouse-over a node; scroll a note into view.}
\end{itemize}
See Supplemental video 1 for a video demonstration of the Explorer. We implemented the CO\textsubscript{2} Explorer as a web application in Javascript with D3.js \cite{d3js} and JQuery library. An online version is accessible at \url{https://ss-18.it.helsinki.fi/health}. 
The prototype source code is available at \url{https://bit.ly/3jKby81}.

\subsection{Use case}
Through identifying interactive entities and enabling entity references to assist insight, we demonstrate how the Explorer supports data/note exploration and annotation in an integrated manner. Initially, the map view shows the global emission data of the latest year. The user notices that the Nordic countries have blue colors indicating lower emission values than the rest of the world. She then clicks on the bar attached to the selected year to view notes as the year relates to significant numbers of notes than other years. Browsing through the notes, the user notices others were comparing emissions of neighboring countries. She then selects Finland and Sweden from the list to view their historical emissions in the line chart and cites a note comparing UK and Ireland to open note editing. She cites the map of the two countries by clicking on the map 
and inputs, ``Finland and Sweden are close to each other and similar in size.'' Mousing over the line chart, she selects the year 1970 and 2000 by clicking on the corresponding vertical reference lines and inputs, ``Around 1970, Sweden emitted twice as much as Finland, whereas, in recent years around 2000, Finland surpassed Sweden.'' The user then publishes the note.

\begin{figure}
    \vspace{-2mm}
  	\includegraphics[width=\linewidth]{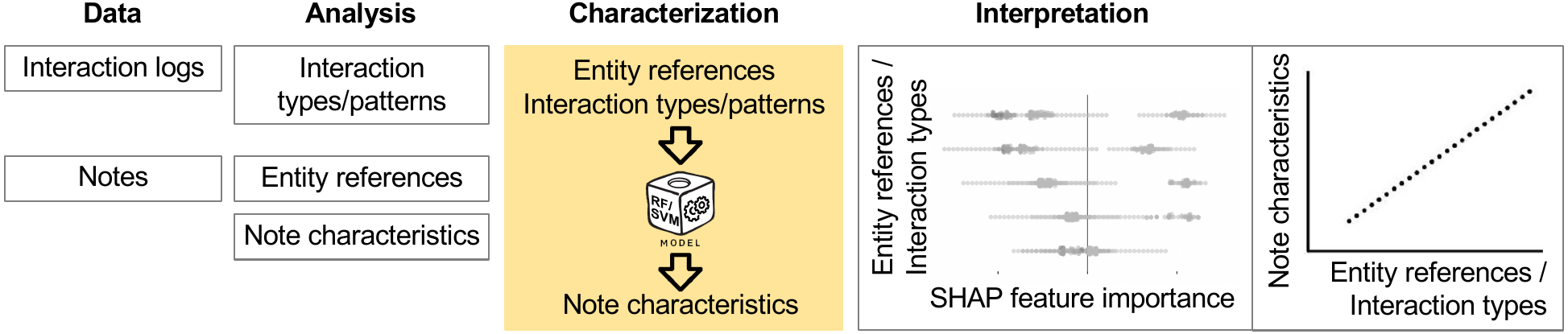}
	\centering
	\vspace{-6mm}
    \caption{The study process. We collected user interactions of using the CO\textsubscript{2} Explorer and their externalized insights in the form of notes and referred entities. We then extracted interaction patterns and characterized notes to prepare for the prediction analysis. See \autoref{sec:results} for a detailed explanation of the characterization using two machine learning models and their interpretation through feature importance and participant-wise correlation analysis.}
    \label{fig:method}
    \vspace{1mm}
\end{figure}

\section{Crowdsourced user study}\label{sec:study}
The study collected entity-based interactions and user insights of exploring the CO\textsubscript{2} emission data, utilized interactions and entity references to characterize notes, and analyzed the prediction performance through feature importance and participant-wise correlations to look into the insight generation process. \autoref{fig:method} provides an overview of the study process.

\subsection{Participants}\label{subsec:particpants}
\begin{figure}
    \vspace{-2mm}
  	\includegraphics[width=\linewidth]{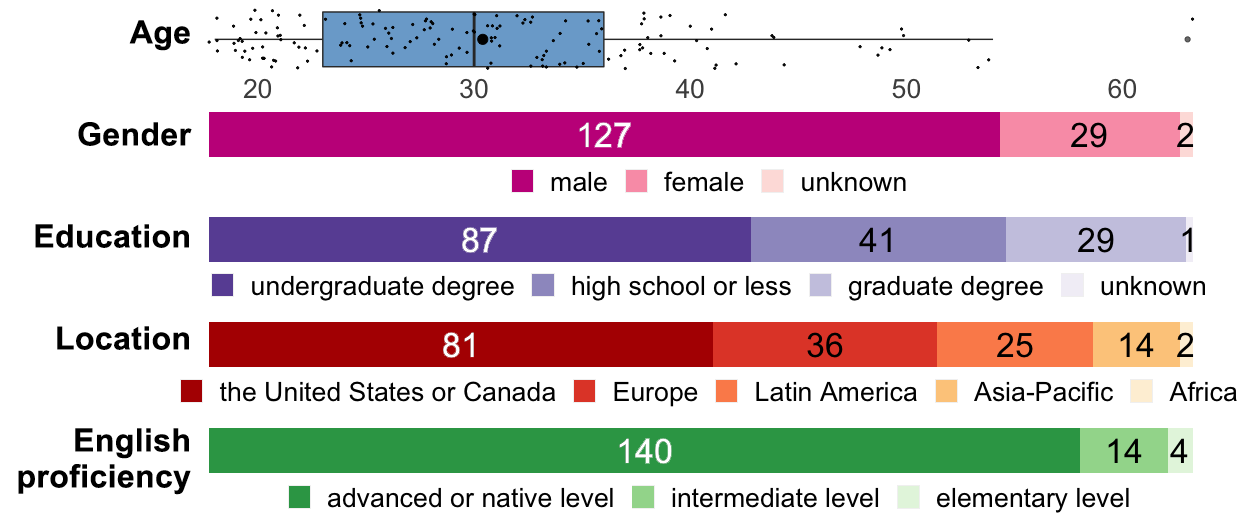}
	\centering
	\vspace{-6mm}
    \caption{A summary of participant demographics.}
    \label{fig:participant}
    \vspace{-2mm}
\end{figure}

We recruited participants through Amazon Mechanical Turk (MTurk), a crowdsourcing platform with workers of diverse backgrounds.
Participants with approval rates of their human intelligent tasks (HITs) greater than 95\% were qualified to perform the task. In total, we approved the HIT submissions of 182 workers. Each received \$1.60 as compensation. We then ruled out six who self-reported as color-blind and 18 who completed the HIT with low quality. For instance, several published notes without data discoveries such as ``India VS China'' and ``note 1;'' two repeated the same sentences in all notes; and one gave neutral answers to all the questions in the questionnaire.  
We used the HIT submissions of the remaining 158 participants for analysis (age range: 18-63, mean: 30.40, SD: 8.92). \autoref{fig:participant} shows the participant demographics regarding their age distribution, gender, education level, location, and English proficiency. Among the participants, the majority (80.38\%) were male; more than half (55.06\%) had an undergraduate degree; around half (51.27\%) were located in the United States or Canada; and the majority (88.61\%) had an advanced or native-level English proficiency.

\begin{table*}[t]
    \centering
    \vspace{-2mm}
    \caption{Note characteristics and assessment criteria.}
    \vspace{-1mm}
\begin{tabularx}{\textwidth}{ >{\raggedleft}p{.2\textwidth} p{.75\textwidth} } 
\toprule
Characteristic & Assessment criteria\\
\midrule
  Category & To categorize notes, we used a bottom-up method and categorized the notes as statements, comparisons, and groupings. We categorized a note as a \textit{statement} if it described a single country, such as \textit{The low rate of CO2 emission of Japan shows that is possible to reach a great level of development sustainably.} We categorized a note as a \textit{comparison} if it compared or contrasted several countries, e.g., \textit{Brazil and Mexico are two emerging countries, whose CO2 rates have been growing in a considerably similar and equal way.} But if the countries were described in a group manner, such as a comparison of Scandinavia countries, then the note was categorized as a \textit{grouping}. For instance, \textit{China leads the Asian countries in CO2 emissions by almost 4 times the next highest country.}\\
  \midrule
  Overview versus detail & We marked 0 if the note described an overview/trend of the data, e.g., \textit{Australia and Brazil have a similar trend.} We marked 1 if the note presented specific values, e.g., \textit{As time goes, in 1971 the CO2 portion of China (831) was approximately 9\% of OECD countries' total CO2 (9,342). However, in 2013, almost 75\% of CO2 emissions from OECD countries (12,037) is from China (8,977).} We marked 0.5 if the note was a mix of the previous two types, for instance, if the note indicated a trend with a year range. For instance, \textit{Russia reduced its emissions from 216,323 to 141,457 in just 7 years and then remained unchanged significantly.} We omitted this aspect if the note did not refer to the data.\\
  \midrule
  Using prior knowledge & We marked 1 if extra information that was not present in the visualization was included in the notes, e.g., \textit{Until 2013 at least the burning of Amazon Rainforest does not place Brazil in the top 5 of CO2 contribution}, and marked 0 otherwise, e.g., \textit{Sweden is reducing their CO2 emissions since 1976 in a very clear way.}\\
  \midrule
  Correctness & We marked 1 if the data description was correct and marked 0 if it was incorrect. We omitted this aspect if the note did not refer to the data.\\
  \midrule
  Relation between notes and referred charts & We marked 1 if the note and the referred chart were relevant and marked 0 if they were irrelevant. We omitted this aspect if no chart was attached to the note.\\
  \midrule
  Entities mentioned & We counted three types of entities (i.e., country, year, and emission value). Besides explicitly mentioned entities, the counts also included derived country, year, and value entities, such as ``Scandinavia countries,'' ``the start of this century,'' and ``increased by 130\%.'' Duplicates did not count.\\
\bottomrule
\end{tabularx}
\vspace{-2mm}
\label{tab:criteria}
\end{table*}

\subsection{Procedure and task}
The MTurk page provided the study link and instruction which described the study as a visualization study that 1) explores how different people make discoveries about the data and 2) would take 30 minutes on average to complete. Following the link was the welcome page which presented the study procedure, system requirements, and a link to the consent form. The participants had to use the Chrome browser on a display at least 1900 pixels wide and 800 pixels high to proceed. By proceeding, they consented to act as research participants.

The study comprised three stages: an interactive tutorial, a visual exploratory task, and a questionnaire. The interactive tutorial guided participants through the main functions of the visualization in 17 steps; it was created using the intro.js library \cite{intro}. 
The steps introduced the map view, the line chart, and then the public notes, and taught users to compose a note through entity references from these views. 
The final introductory step linked the participants to the CO\textsubscript{2} Explorer. After creating an account in the Explorer, the participants proceeded to the task. The task description was as follows:

\vspace{1.5mm}
\textit{Freely explore the data and others' notes, and publish \textbf{at least five notes} which indicate your discoveries about the data. Note: Each note should contain \textbf{at least one type of reference}, such as a view or another note, to help with your narrative. After you finish the task, you can proceed to the questionnaire by clicking on the link at the top bar.}
\vspace{1.5mm}

All participants had access to the same fixed set of public notes. If the participants did not finish the task but clicked on the questionnaire link, a task description window would pop up instead. The questionnaire contained several demographic questions 
and an optional comment space. After the questionnaire, a completion code was provided enabling participants to submit through MTurk to finish the HIT. 

\subsection{Data collection and analysis}\label{sec:dataAna}
We logged entity-based interactions with timestamps and recorded notes and their referred entities. Mouse-overs were recorded if they were longer than 3 seconds. 
In total, we collected 55 entity-based interactions (\autoref{app:type}). 

\subsubsection{Extracting interaction patterns}
Following Guo et al.'s method \cite{caseStudy}, we extracted patterns from interaction trails. We resolved the limitation Guo et al. mentioned by considering the type of action in the patterns rather than the exact number of actions, as the type of actions can reflect user intention better than the number of actions can \cite{caseStudy}. To this end, we considered sequences of the same interactions or at least two different types of interactions as candidate patterns and extracted the patterns in three steps:

First, we counted the consecutive occurrences of the same interactions. Regardless of the length, the same actions in a row were taken as one type of sequence. An exception was that we considered ``deselect a country'' to be the same interaction as ``select a country,'' as they indicated the same intention of exploring countries. Then, we substituted the same consecutive actions by one occurrence in the trail. 

Second, we iterated through the updated trail and recorded the occurrences of all sequences with 2 to 10 interactions. The same action could be counted in multiple sequences to make an exhaustive list of sequences. As an example, consider the sequence of ``select a country $\rightarrow$ open a note input $\rightarrow$ cancel the note input.'' The recorded sequences are ``select a country $\rightarrow$ open a note input,'' ``select a country $\rightarrow$ open a note input $\rightarrow$ cancel the note input,'' and ``open a note input $\rightarrow$ cancel the note input.'' 

The sequences that occurred in more than 39 (25\%) participants' trails were considered candidate patterns. As the candidate patterns from the first iteration counted each action once as we intended, we only considered the ones from the second step in this iteration. We iterated through the updated trail again to match the candidate patterns. This time, one action was counted in at most one candidate pattern. Longer candidate patterns matched first. If there was a match, the next match started from the action next to the matched sequence; otherwise, we iterated to the next action to match again.
The final patterns were the ones that occurred in more than 31 (20\%) participants' trails. We performed this iterative step again to finalize the count of the patterns. 

The two thresholds we mentioned (39 and 31) are adjustable to get various numbers of patterns. However, this adjustment can get tricky. Consider a sequence of ``action 1 $\rightarrow$ action 2 $\rightarrow$ action 3.'' Filtering through the first thresholds of varied values, suppose we can get both ``action 1 $\rightarrow$ action 2'' and ``action 2 $\rightarrow$ action 3'' or ``action 2 $\rightarrow$ action 3'' only as candidate patterns. The two situations result in different counts of candidate patterns to pass through the second threshold. Therefore, we suggest setting the first threshold high enough to filter out trivial candidate patterns and low enough to catch a decent quantity of patterns relative to the number of action types.
\autoref{app:pat} lists 29 interaction patterns extracted from interaction trails. 


\subsubsection{Assessing note characteristics}
As explained in \autoref{relatedWork:chara}, we characterized notes by category, overview versus detail, using prior knowledge, and correctness.
Additionally, tailoring to the insight externalization feature, we evaluated relationships between the notes and referred charts, as well as the number of entities mentioned in the notes. \autoref{tab:criteria} presents the detailed assessment criteria and the example notes from various participants.

One rater first went through the majority of the notes and summarized the detailed assessment criteria. Upon agreement on the criteria, two raters, the authors of the paper, evaluated 50 randomly selected notes independently. The result showed over 90\% consistency between the two raters for all aspects, except for the overview versus detail. After discussion, the two raters refined the overview versus detail assessment criteria by stating that if a trend description were accompanied by a specific year range, then the note would be taken as a mix of overview and detail. Then, the two raters assessed another 30 random notes on this aspect only. The consistency turned out to be over 90\%. One of the raters then evaluated the rest of the notes based on the refined set of criteria (\autoref{tab:criteria}). \autoref{app:note} lists four example notes involving the various aspects of the characteristics and their ratings.

\begin{figure*}
  	\includegraphics[width=\textwidth]{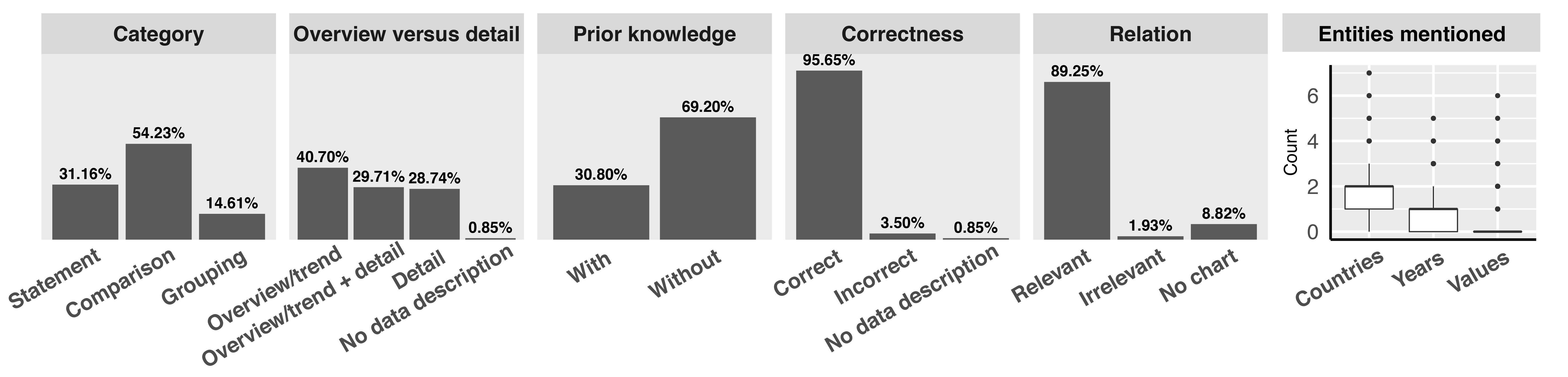}
	\centering
	\vspace{-7mm}
    \caption{Results of the analysis of note characteristics. Except for the last one showing boxplots of entities mentioned in the notes, all others are represented by percentages over the collected 828 notes.}
    \label{fig:quality}
    \vspace{-2mm}
\end{figure*}

\section{Study results} \label{sec:results}
As mentioned in Sec. \ref{subsec:particpants}, after removing unqualified participants, we collected 828 notes from 158 participants. The total number of interactions they performed to complete the task ranged from 41 to 639 (median: 104). 
Specifically, the number of data exploration actions (\autoref{app:type}) ranged from 3 to 541 (median: 50). 
The total time participants spent on the task was between 3 min 56 sec and 1 h 35 min 13 sec (median: 14 min 45 sec). 
We subtracted idle time from the total time when there was no mouse interaction for more than 6 minutes or when the task window was deactivated, indicating the participant might have moved his/her attention to other websites. 

Compared with HindSight \cite{hindsight} and Storytelling \cite{storytelling} which studied a similar CO\textsubscript{2} Explorer, in our experiment, participants performed a comparable number of data exploration actions (HindSight: 54.8, Storytelling: 44) but spent more time (HindSight: 2 min 20 sec, Storytelling: 1 min 49 sec). We attribute the extra time to insight externalization considering the different experimental settings. HindSight evaluated their CO\textsubscript{2} Explorer and collected insights through MTurk as well, but the insight collection stage came after the visual exploration stage, without the participants being able to refer to the visualization, whereas Storytelling studied the visualization in a free-form web-browsing setting without user tasks. Overall, the comparison helped us to confirm the quality of the study.

\textit{Exploration coverage.} 
On average, the participants explored 28 countries (SD: 30) and 9 years (SD: 24), more than in the HindSight study (countries: 7.2, years: 6.7). Different from the HindSight study with a balanced visitation of year and country entities, a Wilcoxon signed-rank test showed that more countries than years were explored in our study (effect size r = 0.64, 95\% CI [0.52, 0.74], p $<$ 0.0001). 

\textit{Note characteristics.} \autoref{fig:quality} presents the evaluation results of note characteristics based on the evaluation criteria described in \autoref{tab:criteria}. The majority of the notes were correct; their referred charts were relevant. This provided a basis for the analysis we performed on the aspects of category, overview versus detail, and using prior knowledge to answer the RQ. For participant-wise correlation analysis, we averaged the assessment results of overview versus detail and using prior knowledge for each participant and counted the note categories of each participant. 
A correlation analysis among these aspects showed that using prior knowledge in notes ($\uptau_b$ = -0.13, 95\% CI [-0.25, -0.02], p = 0.03), as well as the grouping type of notes (-0.13, [-0.25, -0.01], p = 0.04), tended to relate to the overview aspect of notes in the assessment, but the p-values are not statistically significant after Bonferroni correction.

\textit{Entity references.} To gain an overview of entity references, we examined the usage frequency of various types of entities (\autoref{tab:visualEntity}) in collected notes. To do this, we averaged the number of various types of entities referred by each participant in a note and used sign tests to compare the usage of the entity types, as the data violated both the normal distribution and the symmetrical distribution of the differences between variables. We report the median of difference, 95\% CI, and p-value of the sign tests.

\begin{figure}
  	\includegraphics[width=.75\linewidth]{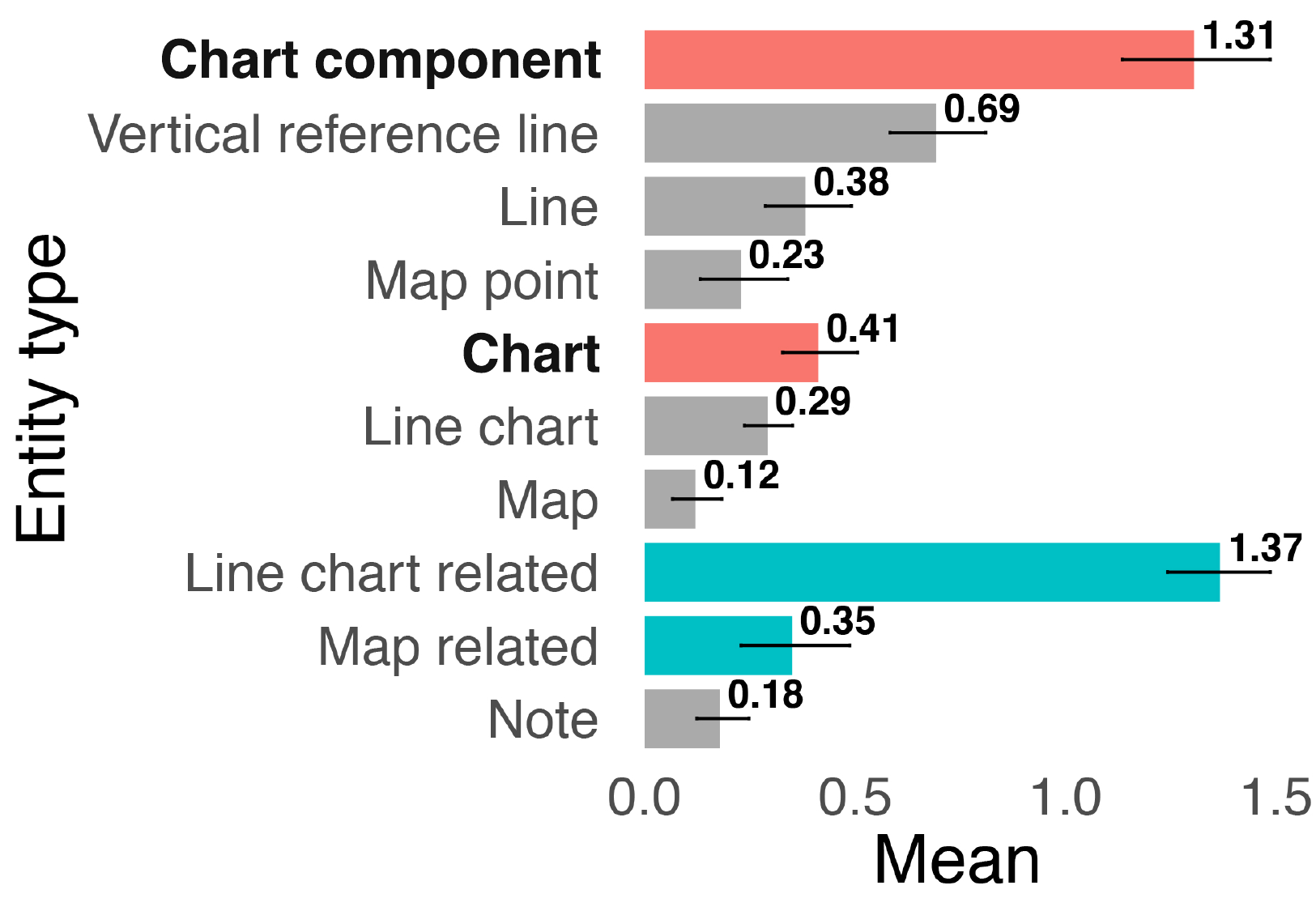}
	\centering
	\vspace{-4mm}
    \caption{The mean and CI of the entities referred to by the participants in a note. CIs are calculated using the percentile bootstrap method with 2,000 replicates.}
    \label{fig:referredEntities}
    \vspace{-3mm}
\end{figure}

\autoref{fig:referredEntities} shows the means and CIs of the numbers of entities referred to by the participants in a note. 
Referring to chart components seemed to have occurred 
more frequently than referring to charts. 
A sign test showed that the difference was statistically significant (median of difference: 1, 95\% CI [0.7, 1.2], p $<$ 0.0001). A series of pairwise comparisons provided evidence that vertical reference lines were referred to more often than lines (0.2, [0, 0.4], p $<$ 0.0001) or line charts (0.4, [0, 0.6], p = 0.0001). The differences between lines and line charts, or map points and maps, were not statistically significant after Bonferroni correction. This might not be a fair comparison, as in visualization, more chart components could be referred to than charts. Nonetheless, the results showed that the various types of references were used for insight externalization, which helps to ground the following analysis. 
Additionally, consistent with the fact that countries were explored more than years, line chart-related references occurred more frequently than did map-related references (1.1, [1.0, 1.2], p $<$ 0.0001).

Next, we elaborate on the analysis we performed to explore the RQ. 
We shared the anonymized study data and R codes of the analysis at \url{https://bit.ly/2QMP4GR} and referred to Dragicevic \cite{fair} when reporting results.

\begingroup
\setlength{\tabcolsep}{4pt}
\begin{table*}[!t]
    \centering
    \caption{Performance of classifying note characteristics using interaction types. Subscripts of S, C, and G under ``category'' represent statement, comparison, and grouping, respectively, whereas the subscripts of O, M, and D under ``overview versus detail'' denote overview, a mix of overview and detail, and detail, respectively, as in \autoref{tab:entityNote}.}
    \vspace{-1mm}
\begin{tabular}{rccrccrcc} 
\toprule
\multicolumn{3}{c}{Category} & \multicolumn{3}{c}{Overview versus detail}& \multicolumn{3}{c}{Using prior knowledge}\\
\cmidrule(lr){1-3}\cmidrule(lr){4-6}\cmidrule(lr){7-9}
 & RF & SVM & & RF & SVM& &RF & SVM\\ 
 Acc. & $0.52 \pm 0.08$ & $0.46 \pm 0.08$ & Acc. & $0.42 \pm 0.08$ & $0.45 \pm 0.08$& Acc. & $0.61 \pm 0.08$ & $0.70 \pm 0.07$\\
  Kappa & $0.22 \pm 0.06$ & $0.15 \pm 0.06$& Kappa & $0.12 \pm 0.05$ & $0.16 \pm 0.06$&Kappa & $0.15 \pm 0.06$ & $0.31 \pm 0.07$\\
 F1\textsubscript{S} & $0.33 \pm 0.07$& $0.45 \pm 0.08$ &F1\textsubscript{O} &$0.51 \pm 0.08$ & $0.56 \pm 0.08$&F1 & $0.68 \pm 0.07$& $0.78 \pm 0.06$\\
  F1\textsubscript{C} & $0.58 \pm 0.08$ & $0.50 \pm 0.08$ &F1\textsubscript{M} & $0.37 \pm 0.08$ & $0.35 \pm 0.07$\\
F1\textsubscript{G} &$0.41 \pm 0.08$  & $0.36 \pm 0.07$ & F1\textsubscript{D} &$0.26 \pm 0.07$  & $0.39 \pm 0.08$\\
\bottomrule
\end{tabular}
\label{tab:actionNote}
    \vspace{2mm}
    \caption{Performance of classifying note characteristics using entity References.}
    \vspace{-1mm}
\begin{tabular}{ r c c  rcc rcc} 
\toprule
\multicolumn{3}{c}{Category} & \multicolumn{3}{c}{Overview versus detail}&\multicolumn{3}{c}{Using prior knowledge}\\
\cmidrule(lr){1-3}\cmidrule(lr){4-6}\cmidrule(lr){7-9}
 & RF & SVM & & RF & SVM&& RF & SVM\\ 
 Acc. &$0.76 \pm 0.07$ & $0.71 \pm 0.07$  & Acc. & $0.52 \pm 0.08$ & $0.51 \pm 0.08$&Acc.&$0.64 \pm 0.08$&$0.68 \pm 0.07$\\
  Kappa &$0.60 \pm 0.08$ & $0.56 \pm 0.08$ & Kappa & $0.27 \pm 0.07$ & $0.26\pm 0.07$&Kappa &$0.23 \pm 0.07$&$0.31 \pm 0.07$\\
  F1\textsubscript{S} & $0.78 \pm 0.06$ & $0.76 \pm 0.07$ & F1\textsubscript{O} & $0.58 \pm 0.08$ & $0.60 \pm 0.08$ & F1 &$0.72 \pm 0.07$ &$0.75 \pm 0.07$\\
  F1\textsubscript{C} & $0.82 \pm 0.06$ & $0.76 \pm 0.07$ & F1\textsubscript{M} & $0.28 \pm 0.07$ & $0.33 \pm 0.07$\\
  F1\textsubscript{G} & $0.43 \pm 0.08$ & $0.54 \pm 0.08$ & F1\textsubscript{D} & $0.56 \pm 0.08$ & $0.54 \pm 0.08$\\
\bottomrule
\end{tabular}
\vspace{-1mm}
\label{tab:entityNote}
\end{table*}
\endgroup


\subsection{Insight characterization} To infer note characteristics, the predictors were the number of various types/patterns of entity-based interactions accumulated until the save/update note action and the number of various reference types of a note. The dependent variables were the note characteristics. 
We utilized two well-known classification models: random forest (RF) \cite{RF} and support vector machines (SVMs) \cite{svm} for the prediction. The ranger \cite{ranger} and randomForest \cite{RFR} R packages were used to tune RF; the e1071 R package \cite{e1071} was used for tuning SVMs. 

We report the accuracy (Acc.), Kappa value, and F1 scores of each prediction. 
We interpreted Kappa values according to Landis and Koch \cite{kappa}, who suggested a value between 0 and 0.20 as slight, 0.21 and 0.40 as fair, 0.41 and 0.60 as moderate, and 0.61 and 0.80 as substantial agreement. For multiclass classifications, the F1 scores were calculated in a one-versus-the-rest manner.
We used 80\% of the collected notes for training and the remaining for testing. The healthcareai R package \cite{healthcareai} was helpful to split data into training and test datasets; the split kept the notes from the same participants in the same dataset while preserving the distribution of the note characteristics across the two datasets.

After removing actions that did not contribute to the prediction and that we considered not to be relevant, we used 35 types of actions to characterize notes. \autoref{app:type} depicts the actions we used as predictors. 
With entity references, we used eight features including number of references to six types of entities (\autoref{fig:referredEntities}), as well as to unique countries and years that were extracted from the referred entities, to characterize notes. 

Tables \ref{tab:actionNote} and \ref{tab:entityNote} show the characterization performance.
We can see that entity references produced better results (i.e., more relevant to note characteristics) than did the interactions that led to the notes. Interaction patterns did not result in better predictions than interaction types, so we do not elaborate on the performance of using interaction patterns. We discuss how to improve insight characterization using interactions in \autoref{subsec:interaction}.

\subsection{Result interpretation}
To gain a more transparent view on the prediction performance, we calculated SHapley Additive exPlanations (SHAP) values \cite{shap} and analyzed the participant-wise correlations of note characteristics to interaction types and entity references. SHAP feature importance measures how much a particular value of a feature changes the outcome of an individual prediction case. The shapper R package \cite{shapper} was used for the calculation.

We used Kendall's tau-b association for the correlation analysis, as the data were not normally distributed. We report the tau-b value, confidence interval (CI), and p-value of the correlation if the absolute tau-b value is over 0.1. We computed 95\% CIs using the percentile method with 2,000 bootstrap replicates. As the correlation analysis was exploratory, not confirmatory, we did not adjust p-values under multiple comparisons \cite{pvalue} but reported the number of analyses we did.

\subsubsection{Using entity references} 

Based on SHAP feature importance, we performed 26 correlation analysis between entity references and note characteristics to interpret prediction results.

\textit{Category.} The number of referred countries was the most important feature in categorizing notes for both RF and SVM. Correlation analysis provided evidence that statements linked to fewer country references ($\uptau_b$ = -0.49, 95\% CI [-0.58, -0.39], p $<$ 0.0001), whereas groupings (0.25, [0.12, 0.37], p $<$ 0.0001) and comparisons (0.29, [0.17, 0.40], p $<$ 0.0001) related to more country references. 
This is consistent with the note assessment that groupings (0.24, [0.11, 0.36], p = 0.0002) and comparisons (0.39, [0.29, 0.49], p $<$ 0.0001) mentioned more countries, whereas statements mentioned fewer (-0.63, [-0.69, -0.55], p $<$ 0.0001). Moreover, groupings appeared to have more chart references (0.28, [0.15, 0.41], p $<$ 0.0001) and fewer chart-component references (-0.13, [-0.26, -0.01], p = 0.04), whereas comparisons showed a weak opposite correlation. That is, comparisons seemed to cite more chart components (0.11, [-0.01, 0.23], p = 0.06) and fewer charts (-0.13, [-0.26, 0.01], p = 0.04), specifically, more citations of the vertical reference lines (0.18, [0.06, 0.30], p = 0.003).

\begin{figure}[H]
    \centering
    \vspace{-1mm}
  	\includegraphics[width=\linewidth]{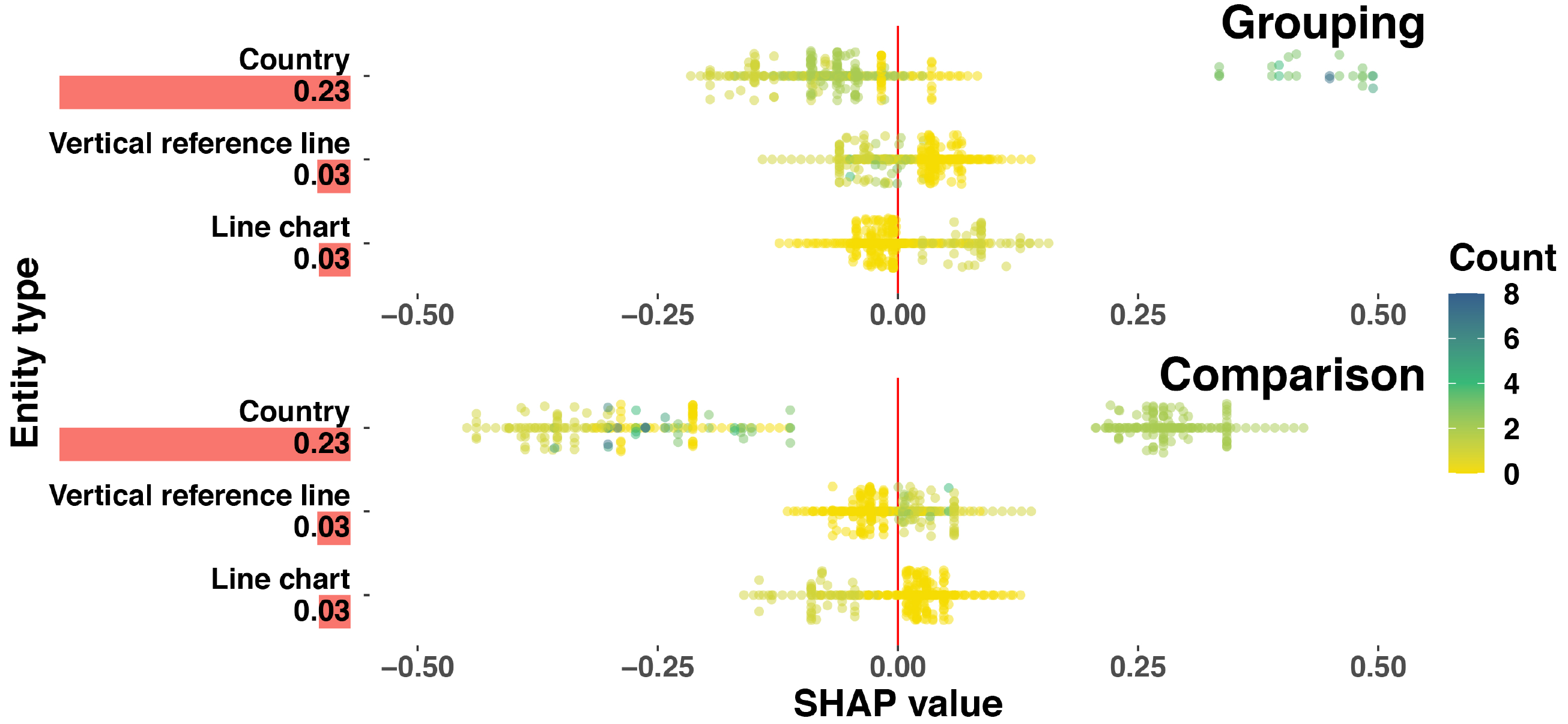}
    \vspace{-6mm}
    \caption{SHAP values of the top three entity types in predicting groupings and comparisons with RF. The vertical axis ranks the top three features with the bars indicating the SHAP values' average absolute deviation; the horizontal axis depicts the features' SHAP values for individual test cases where a negative value lowers the prediction probability, and a positive value does the opposite. A wider spread of SHAP values implies more influence in the model output. Colors represent the original values used in the prediction.}
    \vspace{-1mm}
    \label{fig:cateEntity}
\end{figure}

\autoref{fig:cateEntity} shows the SHAP values of the top three features in predicting groupings and comparisons with RF. Consistent with the correlation analysis, SHAP results showed evidence that high counts of referred countries contributed to the categorization of groupings, whereas counts gathered in the middle around two to four contributed to the comparison category. A large value of vertical reference lines positively affected the categorization of comparisons but had a negative effect on groupings. Conversely, a higher number of line chart references tended to increase the probability of notes being categorized as groupings and lower the note probability as comparisons.

\begin{figure}[H]
    \centering
    \vspace{-3mm}
  	\includegraphics[width=\linewidth]{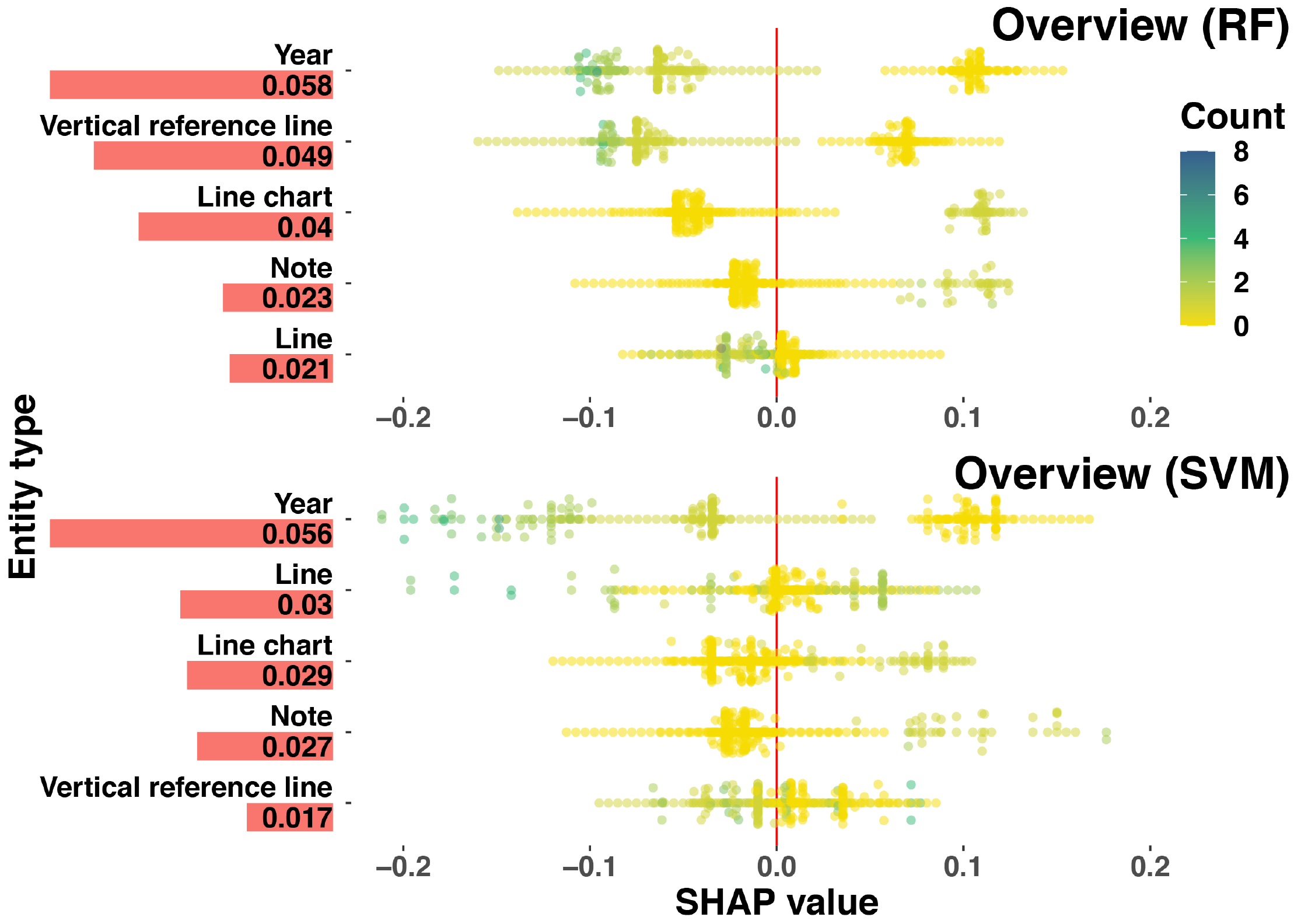}
    \vspace{-6mm}
    \caption{SHAP values of the top five entity types for predicting overviews, the exact reading instructions as in \autoref{fig:cateEntity}.}
    \label{fig:overviewEntity}
\end{figure}

\textit{Overview versus detail.} RF and SVM produced the same set of top five features with a slight difference in the orders (\autoref{fig:overviewEntity}). Overviews seemed to relate to fewer year references and larger numbers of line chart and note references. From the RF chart, greater counts of vertical reference lines and lines tended to lower the probability of overview characterization. Correlation analysis showed a similar tendency that overviews referred to fewer years (0.25, [0.14, 0.37], p $<$ 0.0001), fewer vertical reference lines (0.27, [0.15, 0.37], p $<$ 0.0001), more line charts (-0.24, [-0.36, -0.12], p = 0.0001), and more notes (-0.16, [-0.29, -0.03], p = 0.01).

\begin{figure}[H]
    \centering
    \vspace{1mm}
  	\includegraphics[width=\linewidth]{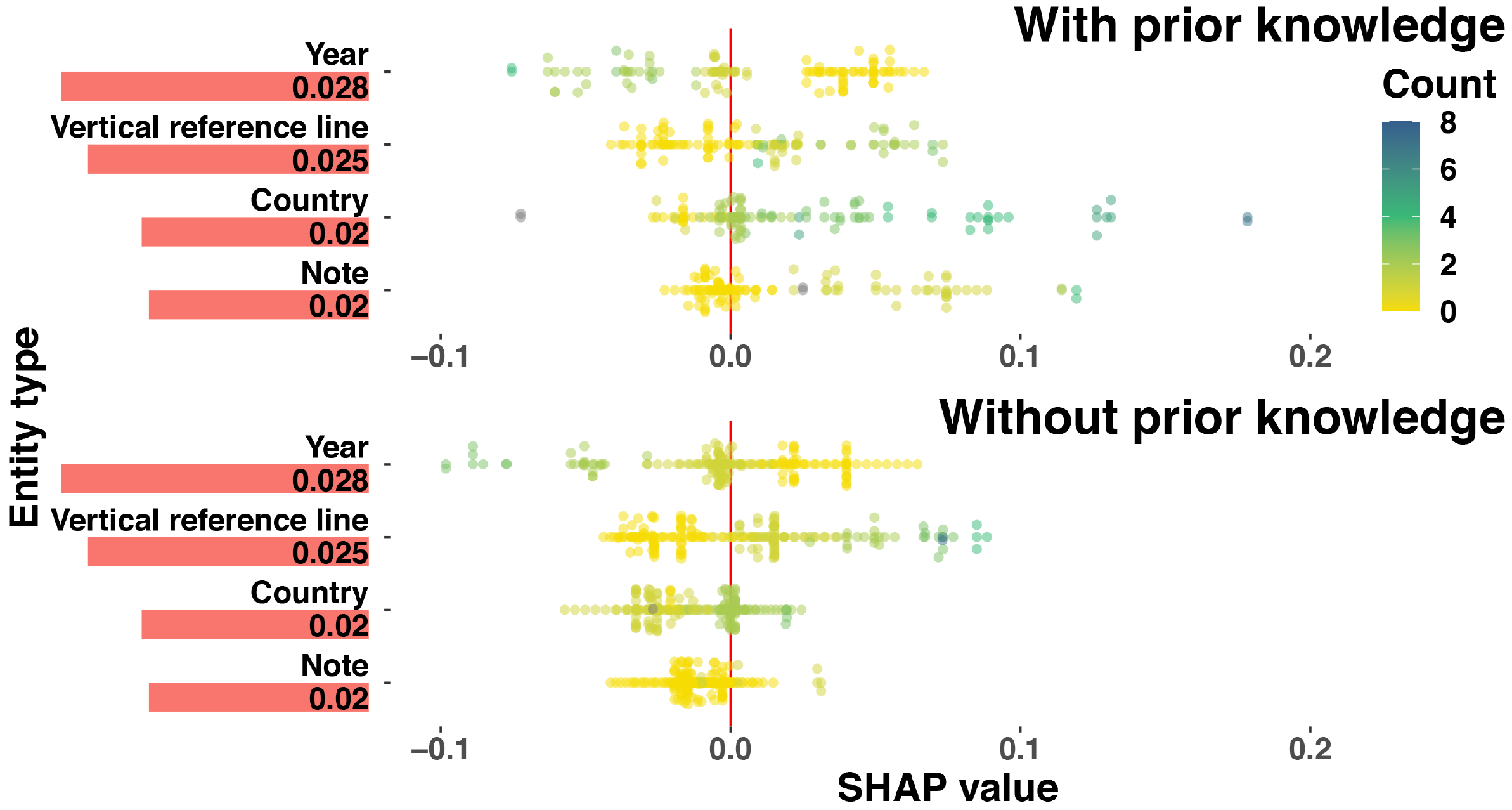}
    \vspace{-6mm}
    \caption{SHAP values of the top four entity types for the predictions of using prior knowledge with SVM. Refer to \autoref{fig:cateEntity} for the instructions.}
    \label{fig:priorEntity}
\end{figure}

\textit{Using prior knowledge.} Based on the SHAP values, the correlation between the features and prediction outcomes seemed uncertain (\autoref{fig:priorEntity}). But there appeared to be tendencies that more country and note references linked to the increased probability of using prior knowledge, whereas larger counts of vertical reference lines made the use of prior knowledge less likely. Correlation results suggested that using prior knowledge positively correlated with note citations (0.15, [0.02, 0.28], p = 0.02). We presume that participants were prone to respond to notes using their prior knowledge. 

\subsubsection{Using entity-based interactions}
Similarly, we calculated SHAP feature importance and the correlations between interaction types and note characteristics to help explain prediction performance. The correlation analysis involved the count of the various interaction types accumulated until the last save/update note action of the participant. To draw a general understanding, we examined interaction types in three groups (data/note exploration and edit actions) as illustrated in \autoref{app:type}, which resulted in 16 analyses.

\begin{figure}[H]
    \centering
    \vspace{-2mm}
  	\includegraphics[width=\linewidth]{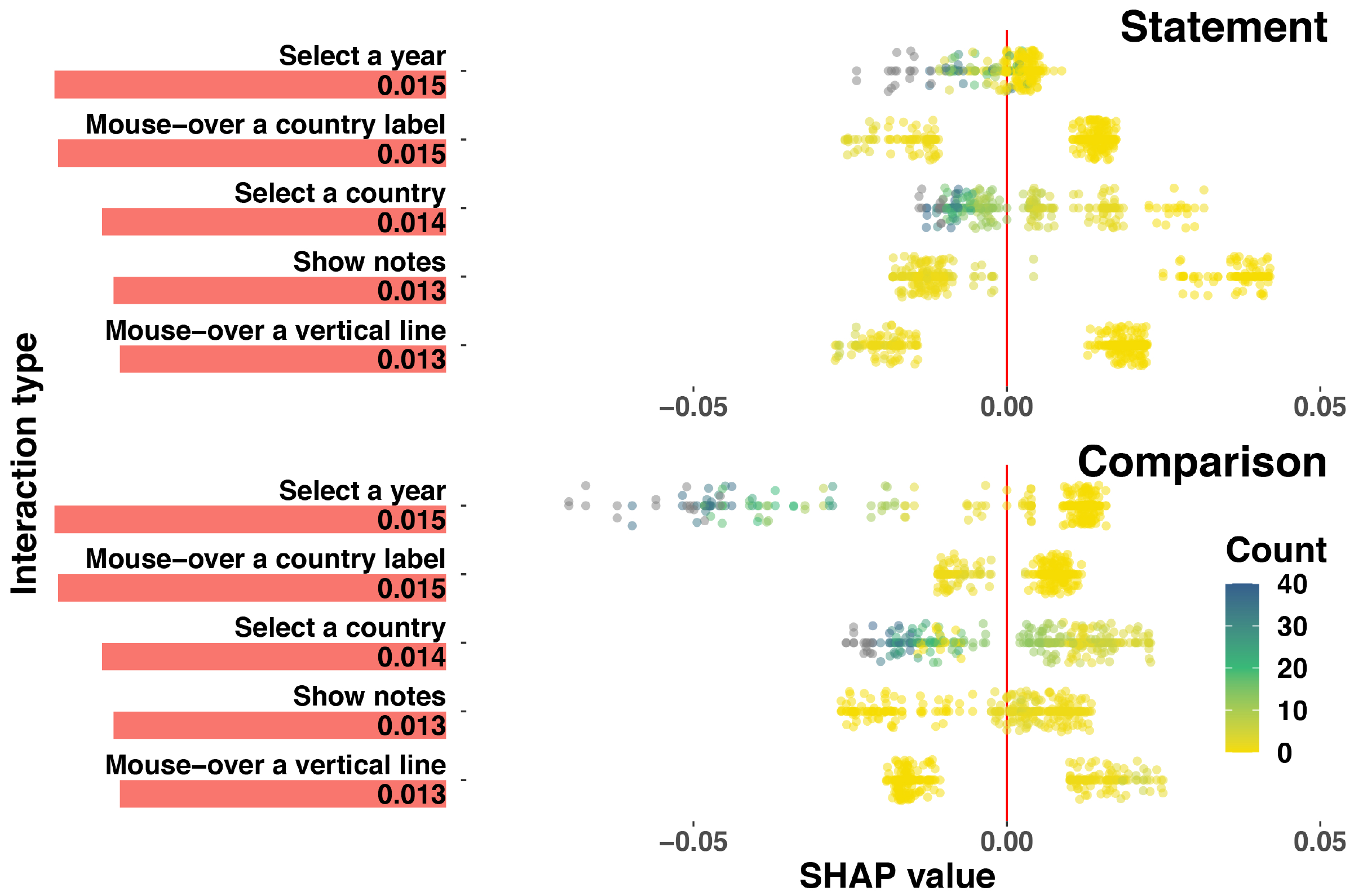}
    \vspace{-6mm}
    \caption{The top five interaction types' SHAP values in predicting statements and comparisons with RF. Refer to \autoref{fig:cateEntity} for the instructions.}
    \label{fig:cateInt}
\end{figure}

\textit{Category.} Correlation analysis showed tendencies that statements related to fewer
data exploration actions ($\uptau_b$ = -0.16, 95\% CI [-0.26, -0.05], p = 0.007); groupings associated to more
data exploration (0.29, [0.18, 0.39], p $<$ 0.0001), more note exploration (0.18, [0.06, 0.29], p = 0.005), and more edit actions (0.13, [0.00, 0.25], p = 0.04); whereas comparisons' link to the actions was uncertain. \autoref{fig:cateInt} shows the SHAP values of the top five features in the cases of predicting statements and comparisons with RF. Compared with predicting statements, large counts of year selections seemed to lower the likelihood of comparisons more; consistent with the earlier finding that comparisons tended to cite more vertical reference lines, mousing-over a vertical reference line appeared to link to comparison categorization.

\begin{figure}[H]
    \centering
    \vspace{2mm}
  	\includegraphics[width=\linewidth]{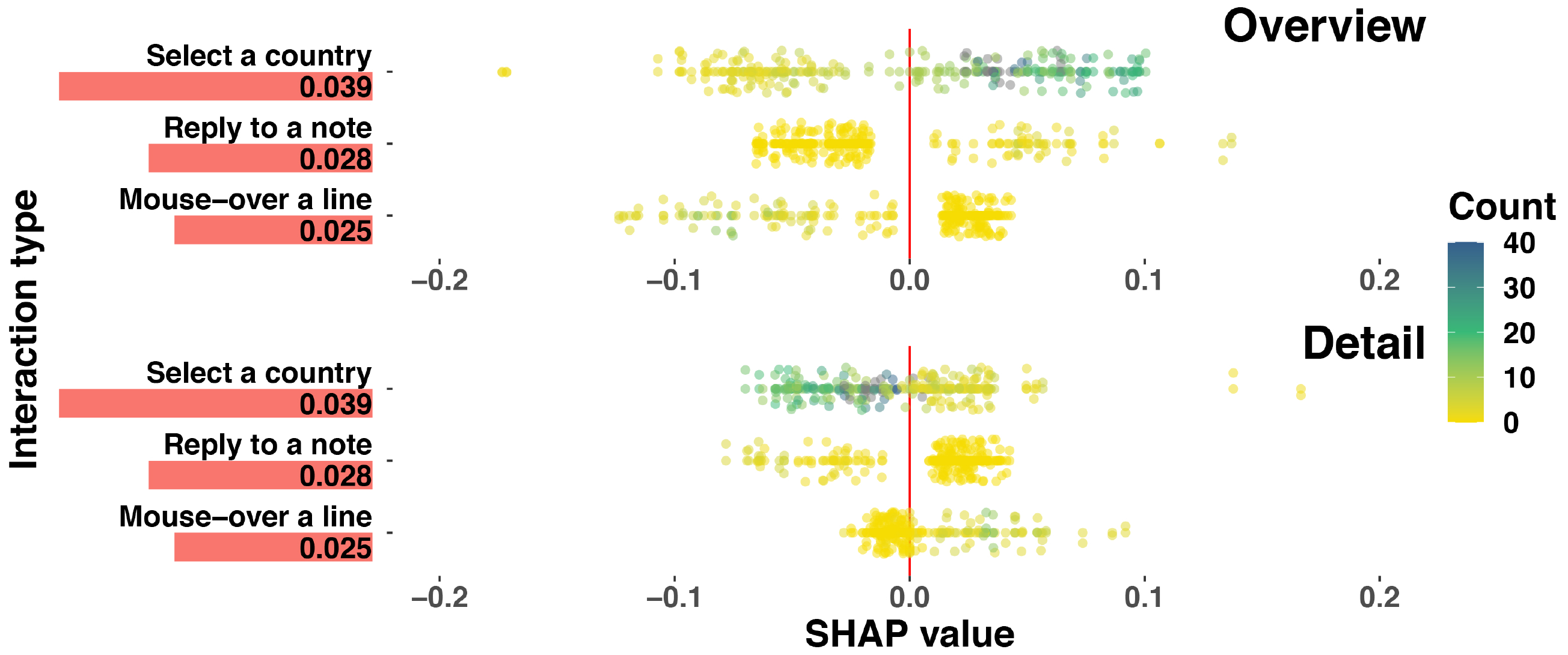}
    \vspace{-6mm}
    \caption{The top three interaction types' SHAP values in predicting overview and detail with RF, the exact reading instructions as in \autoref{fig:cateEntity}.}
    \label{fig:overviewInt}
\end{figure}

\textit{Overview versus detail.} With RF, SHAP values revealed tendencies that details related to fewer country selections and more mouse-overs of the lines in the line chart (\autoref{fig:overviewInt}). Similarly, correlation analysis showed that more detailed notes tended to have more mouse-overs in the map and line chart (0.22, [0.10, 0.33], p = 0.0002); otherwise, there were no significant correlations with the three groups of actions.  

\begin{figure}[H]
    \centering
  	\includegraphics[width=\linewidth]{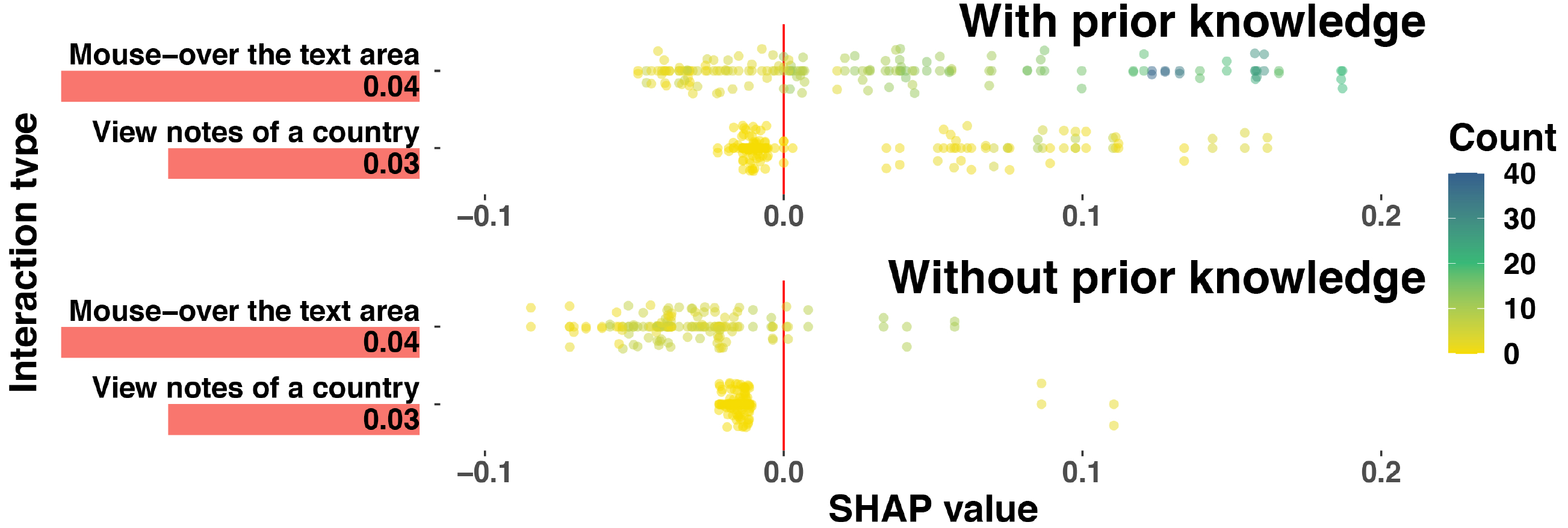}
    \vspace{-6mm}
    \caption{SHAP values of the top two interaction types in inferring the use of prior knowledge with RF. Refer to \autoref{fig:cateEntity} for the reading instructions.}
    \label{fig:priorInt}
\end{figure}

\textit{Using prior knowledge.} Using prior knowledge was inclined to relate to note exploration actions positively (0.13, [0.02, 0.25], p = 0.02), which resonates with the earlier discovery that using prior knowledge tended to have more note citations. \autoref{fig:priorInt} evidences that mousing over the text area of the sticky note and viewing notes of a country largely increased the likelihood of using prior knowledge. 

\section{Discussion and conclusion} \label{sec:conclusion}
To answer the RQ, study results provided evidence that compared with using interactions, using entity references improved note characterization from fair to moderate on categories and from slight to fair on overview versus detail; the two sets of predictors produced comparable fair agreements on using prior knowledge, based on the interpretation of Kappa values \cite{kappa}. Referred entities seemed to be better at predicting insight characteristics than interactions, which is understandable as, during visual exploration, insight is still to be discovered, whereas, during entity references, a resulting insight is made more certain.

SHAP results and participant-wise correlations showed several interesting dependencies. Regarding the three categories, statements seemed to be the simplest, which had fewer data exploration actions and referred to fewer countries, whereas groupings, being more complex, related to more data/note exploration actions and referred to more countries and entire charts. Users were prone to refer to vertical reference lines to assist the narration of details and comparisons.
Notably, more detailed notes tended to have more mouse-overs in the chart area. Using prior knowledge seemed to be linked to more note exploration actions and, consequently, more note references.

Prior work reported both consistent and inconsistent findings to ours. Choe et al. \cite{selfer} had similar results that people referred to chart components to report detail and comparison types of insights. This is not surprising, as we can imagine that references to individual values in the charts afford detailed insight, whereas references to the overall charts afford overview and trend types of insights. However, Choe et al. also collected comparisons against external data, which seemed to refer to whole charts as evidence. 
Regarding prior knowledge, through evaluating a domain-specific visualization, He et al. \cite{he} reported an opposite finding to ours that experts using prior knowledge in visual exploration tended to explore others' notes less. The reason may be with domain-related data, experts were more cautious about shared insights and trusted their own knowledge more, whereas, with generic data, users are more relaxed to respond to other notes using their prior knowledge.

Thus, we conclude that entity-based interactions/references can afford insight characteristics, but the affordances could depend on the types of visualizations; to infer insight characteristics, we need to analyze visualization on a case-by-case basis. For instance, visualization with small multiples may frequently record the comparison type of insight referring to individual charts, different from our finding with the line charts.
However, the research methods we used in this study are generalizable to study other visualizations considering 1) that the insight-based method has been applied with other visualizations in prior work (e.g., \cite{caseStudy, he}), and 2) the generalizability of entity-based interactions, entity references, and insight characteristics in visualization. 
Next, we discuss study implications and limitations.

\subsection{Toward knowledge-assisted visualization}
Knowledge-assisted visualization enables users to externalize and share their knowledge, such as parameter settings and notes, to support visual exploration \cite{dik, explicitKnowledge}. As also discussed by He et al. \cite{he}, the study results imply that characterizing insights automatically through user interactions or referred entities can facilitate personalized insight recommendations. For instance, detailed insights appeared to have more mouse-overs in the chart area; if a user continuously mouse-overs the chart to view detailed data, the tool can proactively recommend detailed insights, as this user is potentially looking for this type of insight. Further, visualization tools can recommend insights with similar or different characteristics to the ones the user has discovered to focus or expand exploration. For instance, if a user discovers insights on comparing countries as interests inferred from referred entities, the tool can recommend comparisons on those countries to the user. Users can also explicitly filter insights by characteristics, i.e., faceted browsing \cite{click2annotate}, while saving the effort and motivation to tag insights manually \cite{taggedComments, commentspace}.

On the other hand, learning from users' entity references, visualization could build insight graphs and guide visual exploration, such as exposing underexplored/overexplored entity relations to nudge users toward serendipitous discovery. Structuring free-form notes by entities and entity relations could further support the process.

\subsection{Limitations}

\subsubsection{Interaction \& Insight} \label{subsec:interaction}
To improve insight characterization using interactions, we need to identify the interactions that are relevant to a specific insight. We collected interactions from the beginning of the task until the save/update note action to characterize notes, even if the captured interactions may be irrelevant to a specific note. However, identifying whether an interaction contributes to a note or not is a non-trivial task. For instance, Zgraggen et al. \cite{multipleComparison} found that interactions that did not result in explicit insight could influence user understanding of the data. We present some potential directions for improvement. Interaction patterns could be extracted based on user goals, incorporating a higher-level semantic segmentation. For instance, researchers proposed methods of segmenting interactions into meaningful chunks considering the action, intent, and data attributes \cite{Tessera} or using topic modeling \cite{topicModeling}, which could potentially benefit insight characterization.
We can also simply take interaction types/patterns that operate on the entities mentioned in insights as effective interactions. In this case, a more detailed level of interaction needs to be used, such as \textit{select the year 2001} instead of \textit{select a year}.

\subsubsection{Entity references}

\textit{Construct the result view directly from references.} Three participants reported being confused about the entity reference mechanism. An improvement can be to construct the result view directly in the sticky note area and enable users to click on the visual elements to remove unwanted references. In the study, the references were shown as snippets of texts explaining attached entities (\autoref{fig:interface} (C)) which may have created an understanding gap in terms of the result view.

\textit{How to support visual exploration and entity references without interfering with one another?} One participant suggested selecting a country by clicking on the map, but we did not support this functionality as it conflicted with the clicks to cite entities. There needs to be more consideration on harmonizing data exploration and entity references, such as by using mouse right-clicks or clicks with an extra key pressed to refer to visual entities or by using double-taps on a touch device. The goal is to prioritize data exploration, implementing insight externalization as an additional interaction layer on top of the data exploration. Additionally, more advanced reference methods can be devised, such as groups of data selection through legends \cite{ChartAccent}, query relaxation \cite{selection}, and range selection in the coordinate system.

\subsubsection{Possible study design-induced biases} 
\textit{Tutorial.} The training tutorial before the actual experiment introduced the Explorer in the order of the map view, the line chart, and the notes, which could affect how users interact with the views according to the studies on visual anchoring biases \cite{anchor, visualAnchor}. It is interesting to see how guidance on a different visual order would impact user interactions. 

\textit{Entity references.} Relating to the anchoring effects, the allowable types of entity references may influence the types of generated insights through entity affordances. For instance, results suggested that vertical reference lines afforded the narration of details. Other types of references, such as data selection through legends, may impact the types of insights users can derive from the charts, which requires further study.

\textit{Task.} The experiment task required users to cite at least one entity in a note, which could introduce bias as well in the collected notes. During the study, participants could access a fixed set of public notes, which could inspire but also impede participants' thinking and, consequently, the diversity of the resulting insights despite the open-ended task.

\section{Supplemental material}
Supplemental video 1 is a video that demonstrates the CO\textsubscript{2} Explorer. (MP4 32,076 kb)

\acknowledgments{We would like to thank Barı\c{s} Serim and the anonymous reviewers for their constructive comments, which greatly helped improve this paper. This work was supported by the Strategic Research Council at the Academy of Finland [Grant Number 335194].}

\bibliographystyle{abbrv}

\bibliography{template}

\clearpage
\appendix
\renewcommand\thefigure{\thesection.\arabic{figure}}    
\setcounter{figure}{0} 
\onecolumn
\section{List of interaction types} \label{app:type} 
We collected 55 interaction types in total (listed below). To make predictions, we removed four actions that strongly correlated with their counter parts such as ``remove discussion'' as a reaction to ``view discussion,'' and three actions that either performed by less than two participants or performed once or twice by all participants (actions \dotuline{under dotted lines}). To classify note characteristics, we further removed several actions that did not contribute to the prediction and that we considered not to be relevant (actions \dashuline{under dashed lines}). In the end, we used 48 action types to predict personality traits, and 35 types to characterize notes.
\begin{itemize}
    \item 12 data exploration actions. Interact with the left country list: \textit{select a country; deselect a country; deselect all selected countries; mouse-over a country.} Interact with the top year list: \textit{select a year; mouse-over a year; play; \dotuline{stop}.} Interact with the map: \textit{mouse-over a map point.} Interact with the line chart: \textit{mouse-over a line; mouse-over the area of a specific year; mouse-over a vertical reference line.}
    \item 22 note exploration actions. Control the note display: \textit{show notes; \dashuline{hide notes}; \dashuline{only show my notes}; \dashuline{also show public notes}.} Scented note browsing: \textit{view notes of a country; view notes of a year; remove the note filter.} Interact with the content of a note: \textit{mouse-over texts of a note; mouse-over a referred note in a note.} Interact with the attached visualization of a note (similar to the data exploration actions): \textit{select a year; mouse-over a year; play; \dotuline{stop}; mouse-over a map point; mouse-over a line; mouse-over the area of a specific year; mouse-over a vertical reference line.} Explore discussions: \textit{view discussions; \dotuline{remove the discussion filter}; drag a node; \dotuline{mouse-over a node}; scroll a note into view.}
\item 14 edit actions. Enter or exit note editing: \textit{\dashuline{open a note input}; reply to a note; \dashuline{save a note}; \dashuline{cancel a note input}; \dashuline{open a note editing}; update a note; \dashuline{delete a note}.}
During editing: \textit{mouse-over the text area; mouse-over a referred entity in the sticky note; add an entity; \dashuline{add an entity repeatedly}; remove an entity; remove a country from the vertical line reference; remove a country from the line chart reference.}
    \item 7 other actions: \textit{\dotuline{Start a session}; \dotuline{de-activate the task window}; \dashuline{activate the task window}; 
\dashuline{check the tutorial}; \dashuline{check the task}; \dashuline{check the questionnaire}; \dotuline{go to the questionnaire}.}
\end{itemize}


\section{List of interaction patterns} \label{app:pat}
Based on the description in \autoref{sec:dataAna}, we extracted 29 patterns from collected interaction trails. The actions in the patterns below can have one or more consecutive occurrences, as the type of action is more important than the exact number of occurrences \cite{caseStudy}.
\begin{itemize}
    \item 5 entity selection patterns interacting with the country or year lists. \textit{select or deselect countries (SelCountry); deselect all selected countries $\rightarrow$ SelCountry; mouse-over a country $\rightarrow$ SelCountry; select years; select a year $\rightarrow$ SelCountry.}
    \item 6 transition patterns among data/note exploration and editing. Data exploration: \textit{SelCountry $\rightarrow$ mouse-over a vertical reference line; SelCountry $\rightarrow$ mouse-over a line; mouse-over a map point $\rightarrow$ SelCountry.} Exploration from data to notes: \textit{SelCountry $\rightarrow$ show notes.} From data exploration to note editing: \textit{mouse-over a vertical reference line $\rightarrow$ open a note input; SelCountry $\rightarrow$ add an entity.}
    \item 14 note editing patterns. Referring entities: \textit{add entities; add an entity repeatedly $\rightarrow$ add an entity; remove an entity $\rightarrow$ add an entity; add an entity $\rightarrow$ update a note.} Editing notes or entities: \textit{mouse-overs in the text area of the sticky note (hoverTextarea); open a note input $\rightarrow$ hoverTextarea/add an entity; open a note input $\rightarrow$ hoverTextarea $\rightarrow$ add an entity; open a note input $\rightarrow$ add an entity $\rightarrow$ hoverTextarea; add an entity $\rightarrow$ hoverTextarea $\rightarrow$ save a note; hoverTextarea/add an entity $\rightarrow$ save a note.} Holistic note editing patterns: \textit{open a note input $\rightarrow$ add an entity $\rightarrow$ hoverTextarea $\rightarrow$ save a note; SelCountry $\rightarrow$ open a note input $\rightarrow$ add an entity $\rightarrow$ hoverTextarea $\rightarrow$ save a note.}
    \item 4 other patterns. \textit{hide notes $\rightarrow$ show notes; show notes $\rightarrow$ only show my notes; only show my notes $\rightarrow$ also show public notes; de-activate $\rightarrow$ activate the task window.}
\end{itemize}
       
            

\section{Example notes and their assessments}\label{app:note}
\begin{figure*}[!h]
\vspace{-2mm}
  	\includegraphics[width=0.99\textwidth]{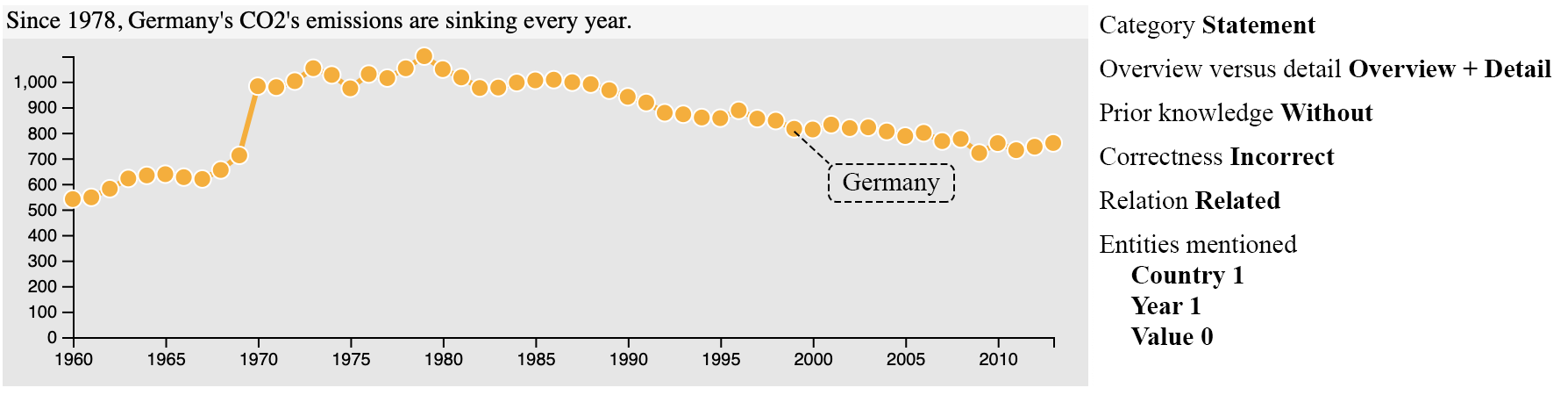}
    \vspace{-5mm}
	\centering
\end{figure*}
\begin{figure*}[!h]
  	\includegraphics[width=0.99\textwidth]{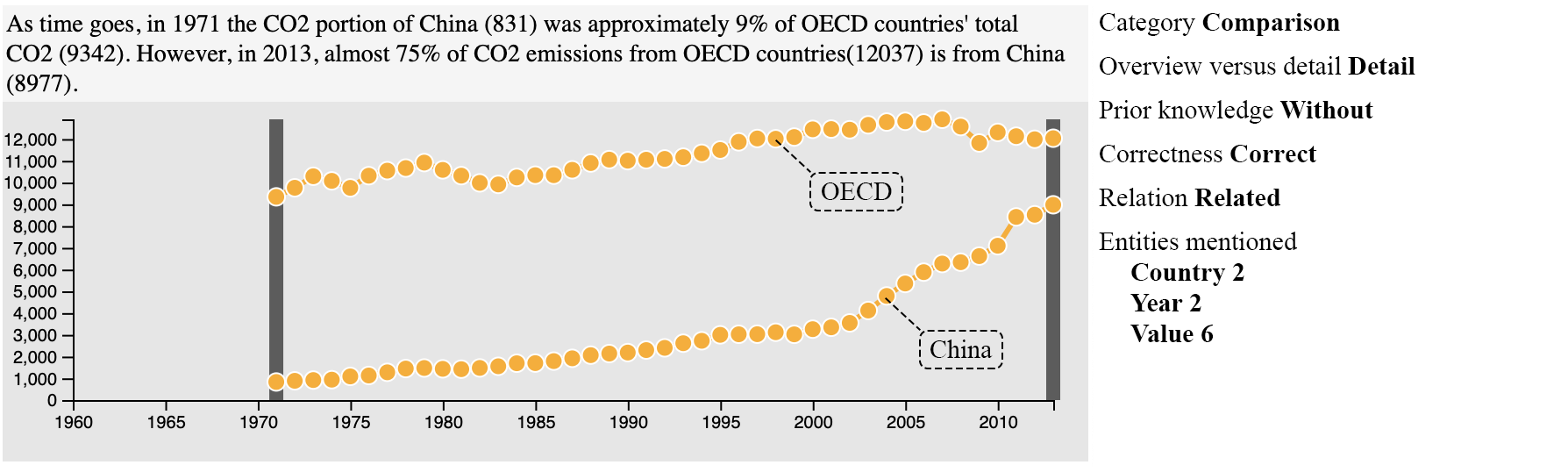}
    \vspace{-5mm}
	\centering
\end{figure*}
\begin{figure*}[!h]
  	\includegraphics[width=0.99\textwidth]{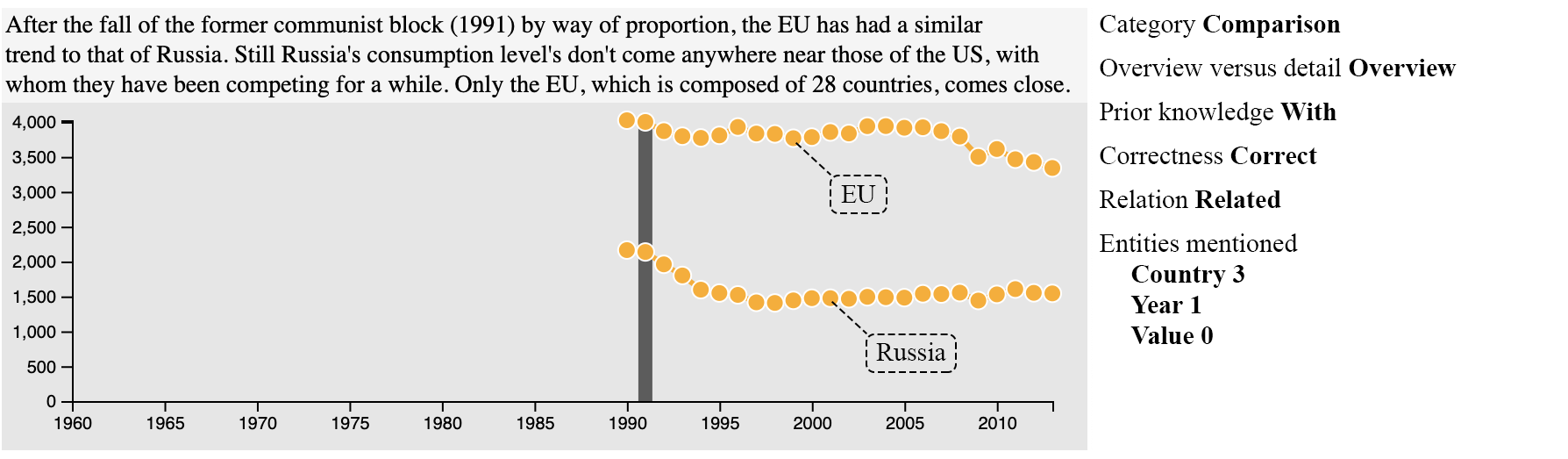}
    \vspace{-5mm}
	\centering
\end{figure*}
\begin{figure*}[!h]
  	\includegraphics[width=0.99\textwidth]{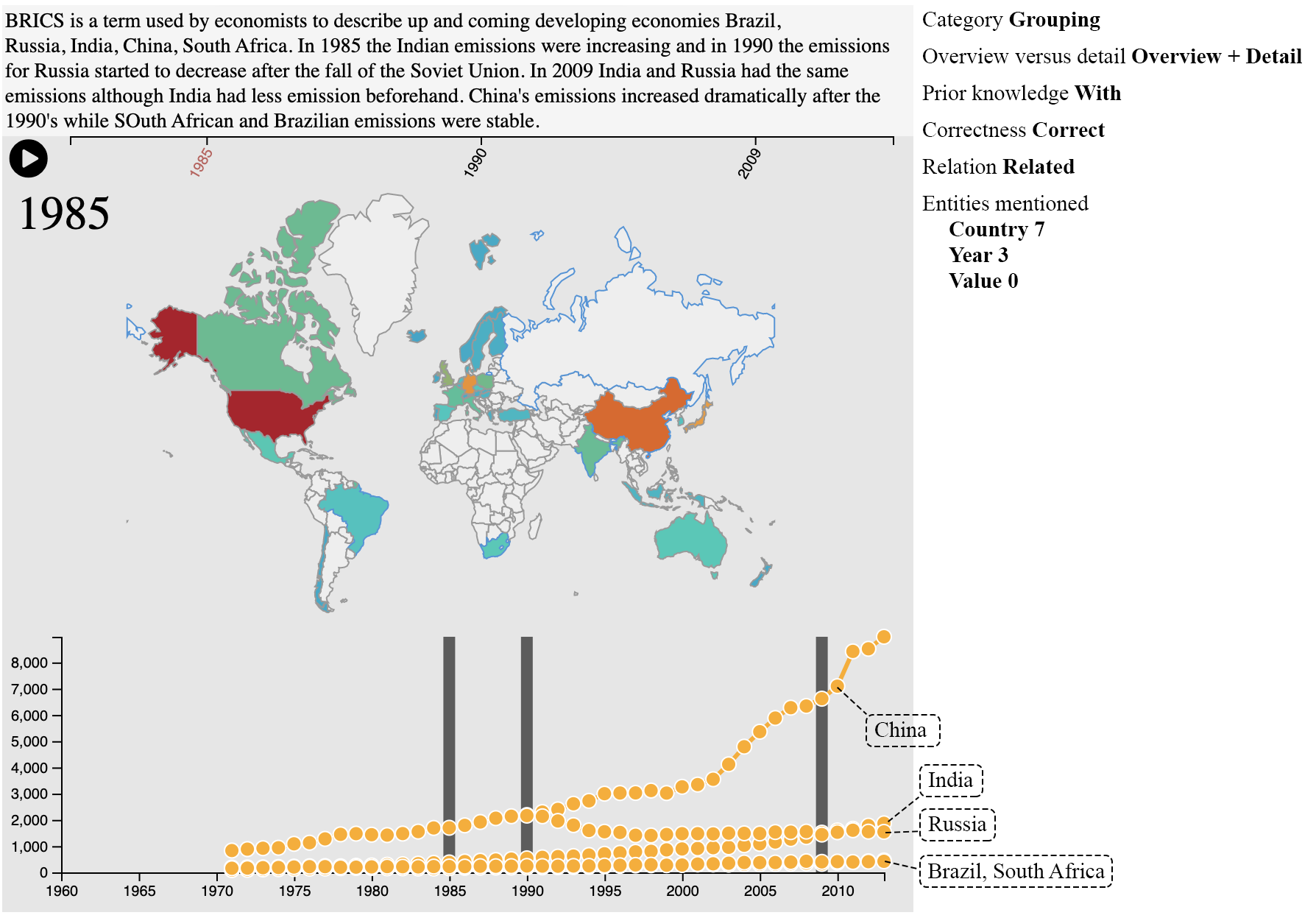}
    \vspace{-5mm}
	\centering
\end{figure*}
\end{document}